\documentclass{article}

\usepackage{arxiv}

\usepackage[utf8]{inputenc}
\usepackage[T1]{fontenc}    
\usepackage{hyperref}       
\usepackage{url}            
\usepackage{booktabs}       
\usepackage{amsfonts}       
\usepackage{nicefrac}       
\usepackage{microtype}      
\usepackage{lipsum}
\usepackage[linesnumbered,lined,boxed]{algorithm2e}
\usepackage{graphicx}
\usepackage[section]{placeins}
\usepackage{subcaption}
\usepackage{wrapfig}
\graphicspath{ {./images/} }
\usepackage{amsmath}

\title{Grocery store flexibility management using model predictive control with neural networks}

\author{
  Roope Sarala \\
  VTT Technical Research Centre of Finland\\
  Oulu, Finland \\
  \texttt{roope.sarala@vtt.fi} \\
   \And
 Jussi Kiljander \\
  VTT Technical Research Centre of Finland\\
  Oulu, Finland \\
  \texttt{jussi.kiljander@vtt.fi} \\
}

\begin{document}
\maketitle

\begin{abstract}
As more and more energy is produced from renewable energy sources (RES), the challenge for balancing production and consumption is being shifted to consumers instead of the power grid. This requires new and intelligent ways of flexibility management at individual building and district levels. To this end, this paper presents a model based optimal control (MPC) algorithm embedded with deep neural network for day-ahead consumption and production forecasting. The algorithm is used to optimize a medium-sized grocery store energy consumption located in Finland. System was tested in a simulation tool utilising real-life power measurements from the grocery store. We report a $8.4\%$ reduction in daily peak loads with flexibility provided by a $20$ kWh battery. On the other hand, a significant benefit was not seen in trying to optimize with respect to the energy spot price. We conclude that our approach is able to significantly reduce peak loads in a grocery store without additional operational costs.
\end{abstract}


\section{Introduction}

Flexibility management is becoming more and more important as increasing amount of electricity is produced from renewable sources. In addition, a large portion of renewable production is close to consumers thus requiring local electricity balancing/management. Large amount of renewable production can cause problems in a traditional grid system not designed local energy production in mind. In order to meet the demands of increasing renewable production, local flexibility management is required at local level. Additionally, flexibility management also provides means for decreasing energy costs, either via shifting load consumption to low spot price periods, or by reducing peak loads that contribute to the energy transmission cost via power tariffs.

Important flexibility providers in a local power grid are grocery stores with large refrigeration systems offering a potential flexibility storage in the form of heat energy. There are various ways of utilizing flexibility in a grocery store, for example,
\begin{itemize}
    \item Load shifting with respect to spot price.
    \item Storing excess production.
    \item Load shifting with respect to peak demand.
\end{itemize}

In recent years, a lot of work has been put in studying building flexibility management with various methods proposed \cite{Shaikh2014b, Chaouachi2013,Thieblemont2017b,Vardakas2015,Vazquez-Canteli2019b}. Generally, these methods can be split into model-free and model-based strategies. As the names suggest, the key difference between these approaches is that the former requires no building model and latter does.

Much of the recent work \cite{Wang2016,Mocanu2019,ONeill2010,Mason2019b} using model-free strategies has been done employing deep neural networks in reinforcement learning framework, such as Q-learning \cite{Sutton1998}. While it is a great approach in flexibility management as it is generally scalable and offers large performance potential, it has some downsides which make them not optimal for every scenario. Mainly, reinforcement learning typically takes a lot of data, obtained via interacting with the physical system, which is not always feasible. In addition, given it is a black-box model, thus the reasoning about its decisions is unknown. This compounded with the fact that it is only able to make good decisions on situations it is familiar with usually means that some additional control logic is needed when controlling critical systems.

Another, more traditional control strategy is model-based predictive control (MPC). It involves a dynamic model of the system, used to predict the future behavior of the system, which is used in optimizing given objective. In addition, a closed-loop or receding approach is taken, where the trajectory is optimized at each time step. Most of the prior work (such as \cite{Ma2012,Privara2013,Freire2008,Candanedo2013a,Hovgaard2013}) use physical models that are robust and data efficient but could suffer from scalability issues and slow computation speeds. In addition, physical models usually require a lot of effort to setup.

In order to combine scalability and performance of the model-free strategies with robustness and data efficiency of MPC, this paper presents a MPC agent fitted with deep neural network model used to forecast day-ahead consumption and photovoltaic production. The agent is used in optimizing supermarket energy costs via reducing peak loads and shifting consumption to low price periods. The performance is validated in a simulation environment using real-world measurements and compared against a rule-based control strategy. In section \ref{sec_methodology}, the optimization problem and the MPC algorithm is described together with the simulation setup. Section \ref{sec_results} is dedicated to presenting the results of both optimization goals with emphasis in peak reduction results. In addition, some control examples is visualized. In section \ref{sec_results}, the results are analyzed and future improvements to the approach is discussed.

\section{Methodology}
\label{sec_methodology}

In this study, a simulated battery component was used as a flexibility resource. This allowed to focus more on the feasibility of the approach instead of intrigues of the refrigeration system dynamics. Subsequently, the control algorithms were tested in a simulation tool built for this purpose with data from real-world power measurements. For forecasting, we used deep neural network models trained with historical data. Optimization was done using gradient-based trust-region method using scikit-learn package for python. We defined cost measurement resolution $T$ as 24 hours and the market resolution $\Delta t$ as 15 minutes, yielding $T = 96$ market steps within one cost resolution.

\subsection{Problem formulation}

The battery can be controlled via three different actions, namely, idle $b^i$, charge $b^c$ and discharge $b^d$. In the following equations, these are used an integers, so that $b^i = 0, b^c = 1, b^d = -1$. The battery output is assumed to be state independent and have equal and constant charging and discharging power $P^{b}$. The optimization period $T$ is divided equally to intervals of time $t$. The optimization problem is then given by the following cost functions, first for peak reduction
\begin{equation}
C_{peak} = max(\sum_{t=1}^{t=T} P_{t}^{s} + b^{i,c,d}_t \cdot P_{t}^{b} ),
\end{equation}
where $P^s$ denotes predicted total non-flexible net consumption at time step $t$. Similarly, for spot price the cost function is
\begin{equation}
C_{spot} = \sum_{t=1}^{t=T} (P_{t}^{s} + b^{i,c,d}_t \cdot P_{t}^{b}) \cdot p_t,
\end{equation}
where $p_t$ is the spot price at time $t$. In addition, we have to take into account physical constraints of the battery, i.e. charge has to stay between $[0,S_{max}]$, where $S_{max}$ is the battery capacity. Expressed in terms of battery charge steps, we have the following linear constraints
\begin{equation}
\begin{aligned}
&\sum_{t=1}^{t=T} b^{i,c,d}_t \geq -s_{t=1}, \\
&\sum_{t=1}^{t=T} b^{i,c,d}_t + s_{t=1} \leq S_{max},
\end{aligned}
\end{equation}
where $s_{t=1}$ is the starting battery level. In addition, as the predictions are never exactly accurate, it is beneficial to keep some charge in the battery in order to respond to a sudden, unexpected changes in the consumption. We can add this to the optimization problem using following additional set of constraints

\begin{equation}
\sum_{t'=1}^{t'=T} \; \sum_{t=1}^{t=t'} b^{i,c,d}_t + s_{t=1} \geq \epsilon,
\end{equation}
where $\epsilon$ is a constant expressing minimum charge amount.

\subsection{Control algorithm}

The action sequence given by the optimization is used to construct a consumption plan. This is done by applying the selected action to the predicted consumption for each time step $t$ so that
\begin{equation}
P_t^{TAR} = P_{t}^{s} + b_{t}^{opt} \cdot P_{t}^{b},
\end{equation}
where $P^{TAR}_t$ is the targe power for timestep $t$. Battery control commands are applied every minute $t_s$, and a closed-loop control algorithm is used to follow the plan by monitoring $\Delta P$, which is given by
\begin{equation}
\Delta P= (\sum_{t_s = 1}^{t_s = T_s} P_{t_s}^{TOT}) - P_{T_s}^{TAR},
\end{equation}
where  $T_s = 15$, the number of minutes in the market resolution. The complete algorithm is presented as pseudocode in algorithm \ref{alg_algorithm}.

\IncMargin{2em}
\begin{algorithm}[ht]
\For{$t_s \in T_s$}{
Forecast consumption and production, i.e. $P^s$\\
Get action sequence $b^{opt}$ from optimizer\\
Recalculate plan $P^{TAR}$\\
Calculate mean realized consumption $P_{t_s}^{TOT}$\\
Calculate difference $\Delta P$ between the plan and current mean\\
\eIf{$|\Delta P| < \phi$}{return $b_i$}{
\eIf{$\Delta P > 0$}{return $b_d$ }{return $b_c$}
}
}
\caption{Control algorithm in pseudocode. $\phi$ is a tunable tolerance parameter between $[0,1]$.}
\label{alg_algorithm}
\end{algorithm}\DecMargin{1em}

\subsection{Data}

\begin{figure}[p]
\includegraphics[width=\textwidth]{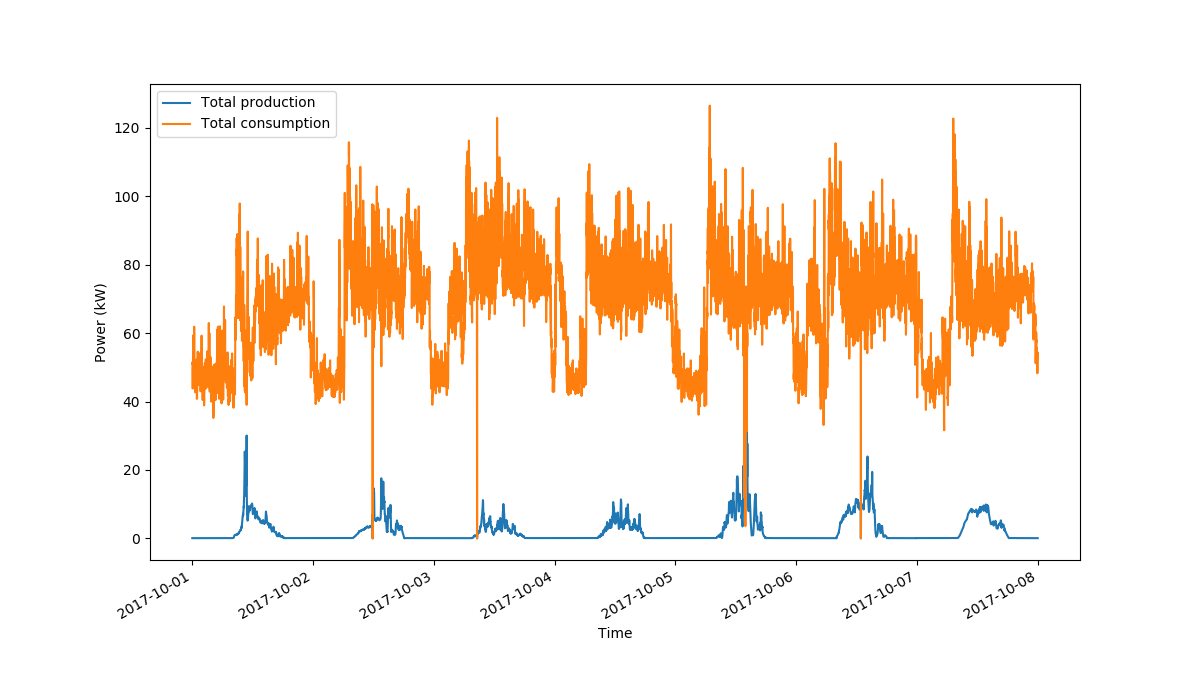}
\caption{Example total consumption of one week. We can see that the consumption and production has a daily cycle that is fairly stable. In addition, total consumption far exceeds production, which is expected during this time of year.}
\label{fig_production}
\end{figure}

\begin{figure}[p]
\includegraphics[width=\textwidth]{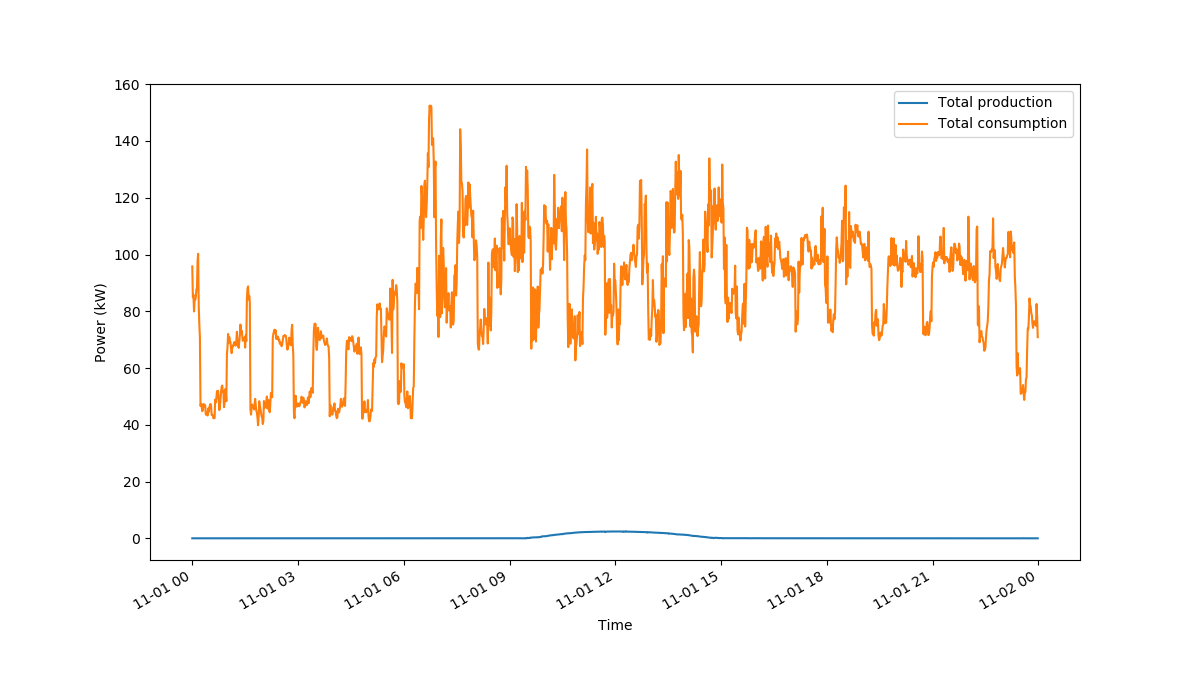}
\caption{Example total consumption of a typical day. Consumption is significantly lower during night time and there is large peak just before the store opens at 07:00 AM. Consumption remains high during the day and then decreases during evening.}
\label{fig_production2}
\end{figure}

The data was collected from a new, medium-sized grocery store fitted with solar panels located in Oulu, Finland from May 2017 to May 2018. Data was measured in one minute intervals from multiple sub-metering points, visualized in figures \ref{fig_production} and \ref{fig_production2}. As we can see from the figures, the main consumption drivers are the refridgerator and heating systems. The electricity spot price was acquired from Nordpool.

\section{Results}
\label{sec_results}

Simulation was run on multiple different battery configurations of varying power and capacity. The default battery power was set as $20$ kW with $20$ kWh capacity. This is equivalent of approximately one hour of capacity with power being 20$\%$ of total consumption.

\subsection{Peak reduction}

\begin{figure}[htbp]
\includegraphics[width=\textwidth]{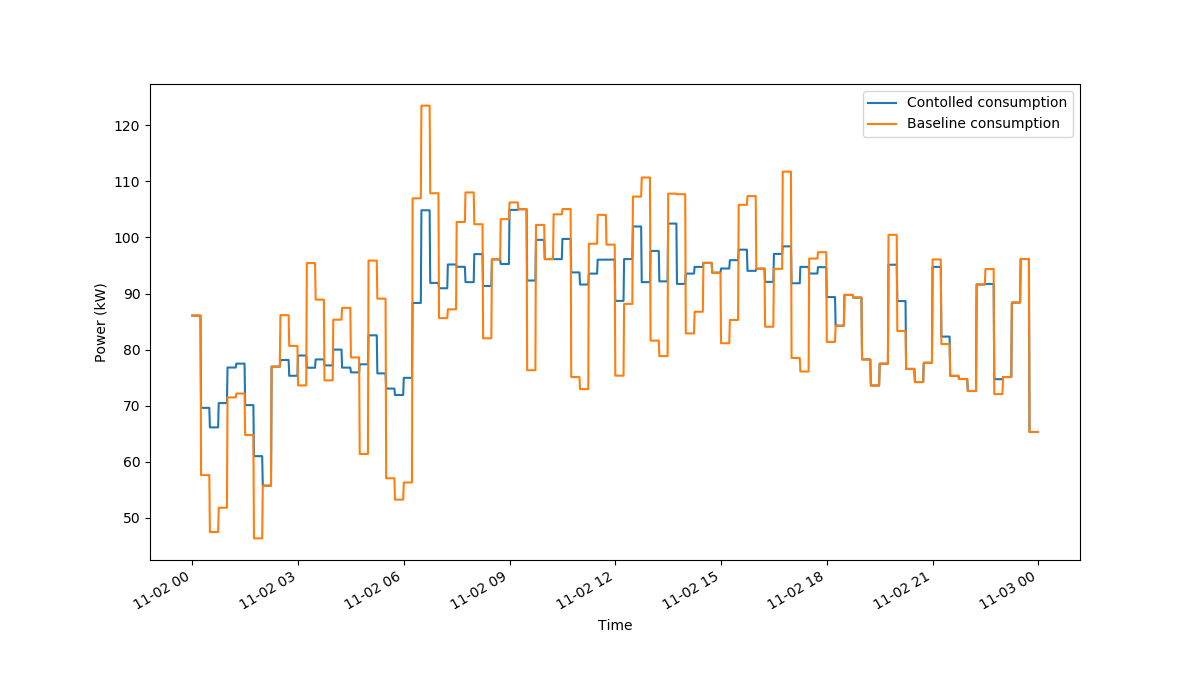}
\caption{Peak reduction example. Blue curve shows controlled consumption whereas orange is the baseline. In this case, the peak was forecasted correctly and algorithm achieved peak reduction of 18\%.}
\label{fig_peak18}
\end{figure}

\begin{figure}[htbp]
\centering
\begin{subfigure}{0.70\textwidth}
\includegraphics[width=0.9\linewidth]{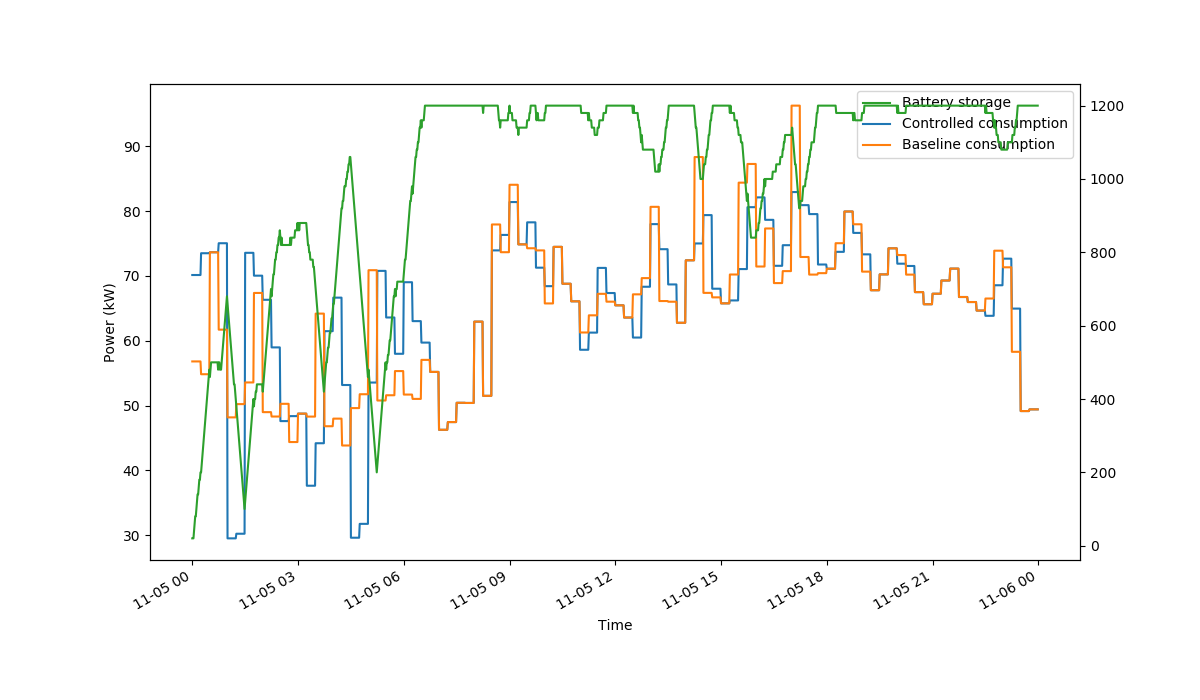}
\caption{Example of a day where there was plenty of capacity left throughout the day and the control algorithm was able to reduce the peak load by $14\%$.}
\label{fig_storage3}
\end{subfigure}

\begin{subfigure}{0.70\textwidth}
\includegraphics[width=0.9\linewidth]{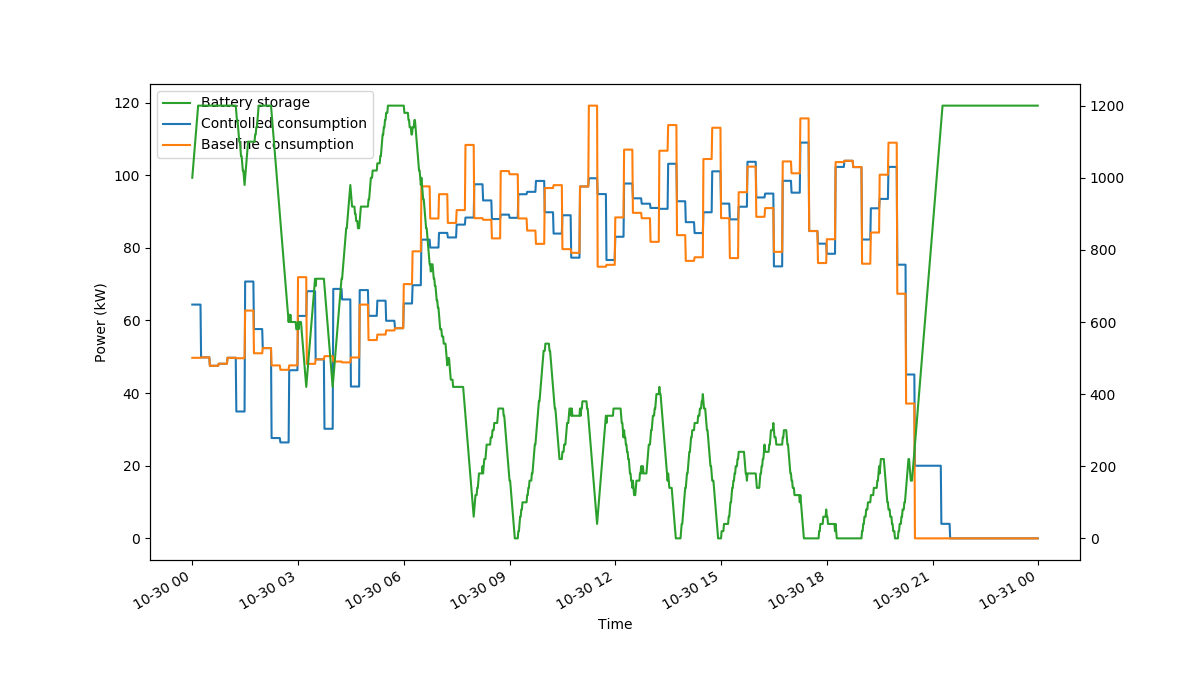} 
\caption{Example of a day where consumption during the day remained high and thus the battery could not be charged. Therefore, control algorithm was not able reduce peak load more than $8\%$.}
\label{fig_storage}
\end{subfigure}

\begin{subfigure}{0.70\textwidth}
\includegraphics[width=0.9\linewidth]{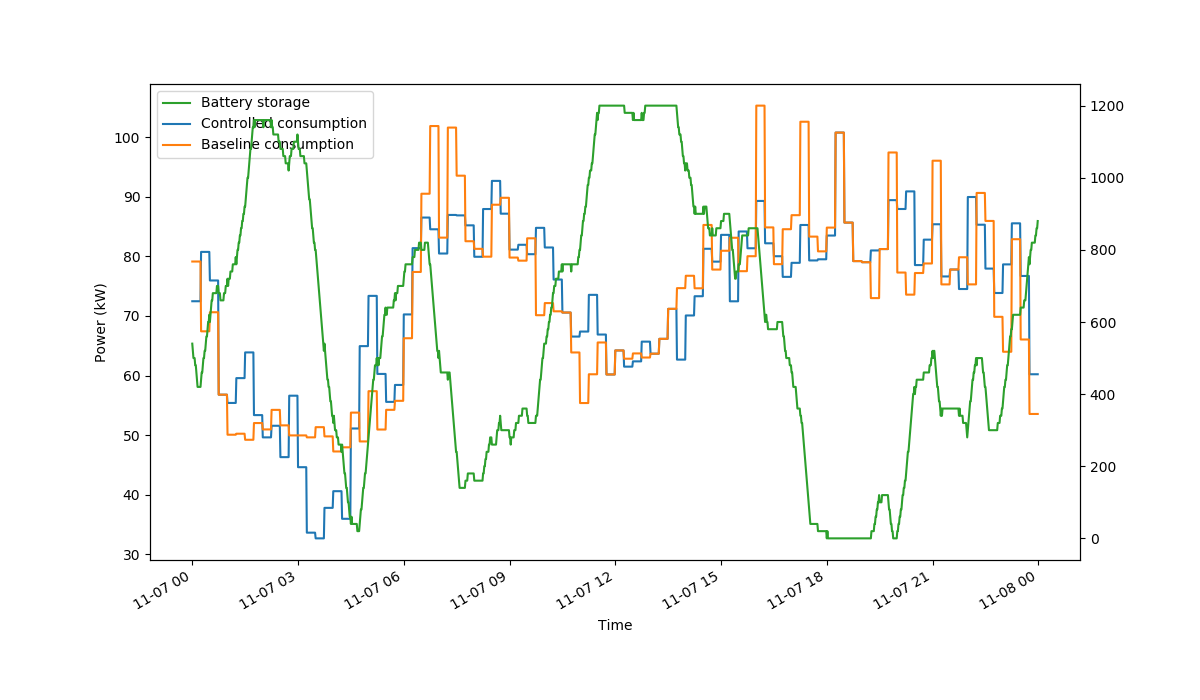}
\caption{Example of a day where multiple high peak loads during evening time discharged the battery and the control algorithm was unable to reduce all peak loads. This resulted in $4\%$ reduction of peak load.}
\label{fig_storage2}
\end{subfigure}
 
\caption{Examples of MPC performance with battery level displayed.}
\label{fig_storages}
\end{figure}

The algorithm performed well when trying to decrease peak consumption (See figure \ref{fig_peak18} for example). With default battery power and capacity, the algorithm managed to reduce peak consumption $8.4\%$ per day. For comparison, with knowledge of future consumption, which can be viewed close to optimal results, the model-based algorithm was able to reduce peak load by $11.2\%$ per day, as seen in table \ref{table_table}. However, the algorithm did perform better than rule-based algorithm by about $13\%$. In addition, results improved with increasing battery power and capacity, as expected. For example, doubling the default battery capacity to $40$ kWh resulted in $9.8\%$ peak reduction for MPC agent and $13.6\%$ with perfect forecasting. Applying a loss of $10\%$ (to $90\%$ efficiency) to the battery efficiency decreased performance around $2\%$. Control examples of different static consumption profiles is presented in figure \ref{fig_storages}.

\begin{wraptable}{r}{0.4\textwidth}
\centering
\begin{tabular}{c|cc}
    Control algorithm & \multicolumn{2}{c}{Optimization objective} \\[5pt]
    & Peak reduction & Spot price \\
    \hline
    Optimal & $11.2 \%$ & $0.6 \%$ \\
    MPC & $8.4\%$ & $0.0\%$ \\
    Rule-based & $7.3\%$ & $0.3\%$ \\[5pt]
\end{tabular}
\caption{Algorithm performance using $20 \text{ kWh} / 20 \text{ kW}$ battery with $100\%$ efficiency.}
\label{table_table}
\end{wraptable}

\subsection{Spot price optimization}

Optimizing with regards to the spot price turned out to be more challenging than expected. With default battery parameters, neither of the algorithms were able to produce any significant savings. Moreover, even with knowing the future consumption, the model-based algorithm managed to decrease costs only $0,6\%$ per day. Performance increased somewhat with increasing battery power and capacity, but not significantly. For example, a cost reduction of only $3\%$ was achieved using $80 \text{ kW} / 160 \text{ kWh}$ battery. Poor performance can be attributed partly to the relative small differences in prices, though it seemed that both large price movements and flexibility storage is needed in order to achieve significant savings.

\section{Discussion}
\label{sec_discussion}

The algorithm was successful in decreasing peak loads, yielding close to $10\%$ reduction. Interestingly, using perfect predictions, peak load only decreased additional couple of percent. This indicates that MPC is not very sensitive to forecasting accuracy. Furthermore, it suggests that this approach may well be applicable in complex systems, where forecasting future consumption is difficult. It was observed that the main limiting factor in the peak reduction performance was the battery capacity. Often the battery would run out of energy in the middle of the day, when consumption was generally high and thus could not reduce the peak optimally. If the battery capacity was larger, thus lasting longer, it could have been more optimally utilized.

With regards to spot price, the approach could not significantly decrease costs in any practical battery configuration. However, increasing battery power and capacity did improve results but not significantly. Poor performance is also attributed to the relatively low variance in energy spot prices.

This study was limited by the lack of actual control data and the subsequent lack of testing in the actual supermarket. The dynamics of the refrigeration system of a grocery store are more complicated than that of a battery, thus to further evaluate the success of this approach, a real-life control experiments would be preferable. In addition, the algorithm could improved to include reinforcement learning to control various tunable parameters or more sophisticated forecasting setup could be used.

\bibliographystyle{unsrt}
\bibliography{main}

\end{document}